# A Sparse PCA Approach to Clustering[1]

T. Tony Cai and Linjun Zhang

Department of Statistics

The Wharton School

University of Pennsylvania

February 16, 2016

## 1 Introduction

Clustering of high-dimensional data under the Gaussian mixture model is an important problem in statistics. In the high-dimensional setting, classical clustering methods, such as the Expectation-Maximization (EM) algorithm, do not perform well due to the large number of free parameters. [6] proposes a two-stage IF-PCA procedure for clustering high-dimensional Gaussian mixtures with a diagonal covariance matrix. The model with cluster size $K$ is assumed to be of the following form:

$$X_i \overset{i.i.d.}{\sim} \sum_{k=1}^{K} \delta_k N(\mu_k + \bar{\mu}, \Sigma),$$

where $\bar{\mu} = \frac{1}{n}\sum_{i=1}^{n} \mathbb{E}[X_i]$, and $\mathbb{E}[X_i] = \bar{\mu} + \mu_k$ if $X_i$ belongs to the class $k$. $\Sigma \in \mathbb{R}^{p \times p}$ is assumed to be diagonal, and $\mu_1, ..., \mu_K$ are assumed to be jointly sparse. Formally, let $M$ be a $K \times p$ matrix with $k$-th row being $\mu_k$, and we assume $\max_j ||M_{j*}||_0 \leq s$.

One can also write the observed data matrix $X \in \mathbb{R}^{n \times p}$ as

$$X = 1_n \bar{\mu}^\top + LM + Z, \qquad (1)$$

where $L \in \mathbb{R}^{n \times K}$ with its $i$-th row being $e_k$ if $X_i$ belongs to the class $k$. In addition, $Z \in \mathbb{R}^{n \times p}$ has rows $Z_i$ and $Z_i \overset{i.i.d.}{\sim} N_p(0, \Sigma)$.

---

[1] The research of Tony Cai was supported in part by NSF Grants DMS-1208982 and DMS-1403708, and NIH Grant R01 CA127334.

The two-stage procedure proposed in [6] first selects features based on the Kolmogorov-Smirnov statistics and then applies a spectral clustering method to the post-selected data. A rigorous theoretical analysis for the clustering error is given and the results are supported by a competitive performance in numerical studies.

The following comments are divided into two parts. We will discuss a clustering method based on the sparse principal component analysis (SPCA) method proposed in [4] under mild conditions and compare it with the proposed IF-PCA method. We then discuss the dependent case where the covariance matrix $\Sigma$ is not necessarily diagonal. To be consistent, we will follow the same notations used in [6].

## 2 A Clustering Method Based on the SPCA Procedure Given in [4]

In Section 1.6 of [6], the authors showed numerically that the proposed IF-PCA method outperforms a clustering method using the SPCA algorithm introduced in [9]. However, the SPCA method in [9] is not designed for the optimal control of principal subspace estimation error and thus does not perform well in the subsequent clustering. The problem of SPCA has been actively studied recently and several rate-optimal procedures for estimating the principal components and principal subspaces have been proposed. See, for example, [2, 4, 5, 7].

In this section, we first introduce a clustering algorithm in the setting considered in [6] using the SPCA procedure introduced in [4], which was shown to be rate-optimal for estimating the principal subspace under a joint sparsity assumption. We then make a comparison of the performance of this SPCA clustering procedure with that of the proposed IF-PCA method both theoretically and numerically. The results show that this SPCA based clustering procedure yields a comparable bound for clustering error rate with that of IF-PCA under mild assumptions and it also performs well numerically.

Throughout this section we assume that the common covariance matrix $\Sigma$ is diagonal and $K$ is of constant order. Recall that the normalized data matrix $W$ can be decomposed as:

$$W = [LM + Z\Sigma^{-1/2}]\Lambda + R = UDV^\top + Z\Sigma^{-1/2}\Lambda + R, \qquad (2)$$

where $Z\Sigma^{-1/2} \in \mathbb{R}^{n\times p}$ has *i.i.d.* $N(0,1)$ entries, $L \in \mathbb{R}^{n\times K}, M \in \mathbb{R}^{K\times p}$, and $LM$ is a matrix where the $i$-th row is $\mu_k$ if and only if sample $i \in$ Class $k$. In addition, $UDV^\top$ is the singular value decomposition (SVD) of $LM\Lambda$, with $D \in \mathbb{R}^{(K-1)\times(K-1)}$, and we assume $\lambda \leq \lambda_{K-1}(D) \leq \lambda_1(D) \leq c\lambda$ for some $\lambda$ and constant $c$. In addition $R$ is a negligible term defined in (2.7), and $\Lambda$, which is given in (2.7) of the paper, is a diagonal and approximately identity matrix with $||\Lambda||_2 \leq 1$. Note that $\mu_k$ ($k = 1, ..., K$) are jointly sparse according to the assumptions in the [6], and $\mu_1, ...\mu_{K-1}$ are linearly independent. These imply $V \in G(s, p, K-1)$, where $G(s, p, K-1) = \{V \in O(p, K-1) : ||V||_w \leq s\}$ with $||V||_w := \max ||V_{*j}||_0$, and $O(p, r)$ denotes the set of all $p$ by $r$ matrix with orthonormal columns.

The above discussion shows the connection between (2) considered in [6] and the sparse PCA model studied in the literature. For the sparse PCA model, a reduction scheme was proposed in [4] for estimating the principal subspace span($V$) by transforming the original problem to a high-dimensional multivariate linear regression problem with the orthogonal design and group sparsity. The estimator $\hat{V} \in \mathbb{R}^{p \times K-1}$ is fully data-driven and can be computed efficiently, and is proved to be adaptively minimax rate-optimal. Once $\hat{V}$ is available, the principal subspace span($U$) can be well estimated and applying $k$-means to the estimator $\hat{U}$ leads to a clustering procedure. The following Algorithm 1 formalizes the procedure outlined above by providing the detailed steps of the SPCA method introduced in [4].

---
**Algorithm 1** SPCA clustering method
---
**Input:** The normalized data matrix $W$, parameters $\alpha, \beta, \delta > 0$.
**Output:** The estimated class labels $\hat{y}^{\text{SPCA}} \in \{1, 2, ..., K\}^n$.
1: Generate an independent $n \times p$ matrix $\tilde{Z}$ with i.i.d. N(0,1) entries, and form two samples $W^i = W + (-1)^i \tilde{Z}$ ($i = 0, 1$).
2: Use the sample $W^0$ to compute an initial estimator $\hat{V}_0 \in \mathbb{R}^{p \times (K-1)}$, where the procedure is discussed in Theorem 2.
3: Form $B = W^0 \hat{V}_0$ and let its SVD be $B = QCR^\top$ with $Q \in \mathbb{R}^{n \times (K-1)}$ and $C \in \mathbb{R}^{(K-1) \times (K-1)}$, then let $Y = \frac{1}{\sqrt{2}}(W^1)^\top W^0 \hat{V}_0 R C^{-1}$.
4: Define pen($\Theta$) = pen($|\text{supp}(\Theta)|$), where supp($\Theta$) is the index of nonzero rows of $\Theta$, and pen($k$) = $(1 + \delta)^2 \sum_{i=1}^{k} t_i$ with $t_i = K - 1 + \sqrt{2(K-1)\beta \log \frac{ep}{i}} + \beta \log \frac{ep}{i}$.
5: Let $\hat{\Theta} = \arg\max_{\Theta \in \mathbb{R}^{p \times K-1}} ||Y - \Theta||_F^2 + \text{pen}(\Theta)$.
6: Construct $\hat{V}$ by orthonormalizing the columns of $\hat{\Theta}$.
7: Construct $\hat{U} \in \mathbb{R}^{n \times (K-1)}$ by orthonormalizing the columns of $W\hat{V}$.
8: $\hat{y}^{\text{SPCA}}$ is constructed by performing the $k$-means to the rows of $\hat{U}$, assuming there are $K - 1$ clusters.
---

The estimation error of span($\hat{U}$) and the clustering error of the resulting clustering procedure can be well bounded. The theoretical results are summarized in the following theorem.

**Theorem 1** *Let $\hat{y}^{\text{SPCA}}$ be the estimated label vector obtained by Algorithm 1, with the initial estimator $\hat{V}_0$ satisfying $\sigma_r(\hat{V}_0^\top V) \geq 1/2$, and $|\text{supp}(\hat{V}_0)| \leq s'$, with $s' = p^{1-v} + p^{1-q} +$* $\log p$. *Under the conditions of Theorem 2.2 in [6], suppose $\rho_2(L, M) \lesssim ||\kappa||^2$, $p^{1-v} \log p \lesssim n$, and $\lambda \leq \lambda_{K-1}(D) \leq \lambda_1(D) \leq c\lambda$ for some $\lambda \gtrsim \sqrt{\frac{\log p}{n}}$ and constant $c$, then the clustering error rate of $\hat{y}^{\text{SPCA}}$ satisfies*

$$\mathbb{E}[\frac{1}{n}\text{Hamm}_p^*(\hat{y}^{\text{SPCA}}, y)] \leq L_p err_p,$$

where as in [6],

$$err_p = \rho_2(L, M)[\frac{1 + \sqrt{\frac{p^{1-v \wedge q}}{n}}}{||\kappa||} + p^{-\frac{(\sqrt{r} - \sqrt{q})_+^2}{2K}} + \sqrt{p^{v-1} + \frac{p^{(v-q)_+}}{n}}\sqrt{\rho_1(L, M)}],$$

and

$$\text{Hamm}_p^*(\hat{y}^{\text{SPCA}}, y) = \min_\pi \{\sum_{i=1}^n I(\hat{y}^{\text{SPCA}} \neq \pi(y_i))\},$$

*with $\pi$ being any permutation of $\{1, 2, ..., K\}$.*

*Proof:* Using an analogous argument to the proof of Theorem 6 in [4], together with (C.70) in [6], $\hat{\Theta}$ in Step 5 of Algorithm 1 satisfies

$$\mathbb{E}[||\hat{\Theta} - \Theta||_F^2] \leq Ks' + s' \log \frac{ep}{s'} + p^{(v-q)_+}||\kappa||^2 \rho_1(L, M),$$

and consequently there exists $H \in O(K - 1, K - 1)$, such that

$$\mathbb{E}[||\hat{V} - VH||_F^2] \lesssim \left(\frac{\lambda + 1}{n\lambda^2}\left(Ks' + s' \log \frac{ep}{s'} + p^{(v-q)_+}||\kappa||^2 \rho_1(L, M)\right)\right) \wedge K.$$

The left singular vectors $U \in \mathbb{R}^{n \times (K-1)}$ in Step 7 of Algorithm 1 is estimated by orthonomalizing the columns of $W\hat{V}$. Since

$$\begin{aligned} W\hat{V} &= (UDV^\top + Z + 1_n(\bar{\mu} - \bar{X})^\top)\hat{V} \\ &= UDH + UD(V^\top\hat{V} - H) + (Z + 1_n(\bar{\mu} - \bar{X})^\top)\hat{V}, \end{aligned}$$

it then follows from Wedin's sin-theta Theorem that there is $\tilde{H} \in O(K - 1, K - 1)$ satisfying

$$\begin{aligned} \mathbb{E}[||\hat{U} - U\tilde{H}||_F] &\leq \sqrt{K}\,\mathbb{E}[\frac{||UD(V^\top\hat{V} - H) + (Z + 1_n(\bar{\mu} - \bar{X})^\top)\hat{V}||_2}{\lambda}] \\ &\leq \sqrt{K}\,\mathbb{E}[\frac{||UD(V^\top\hat{V} - H)||_2 + ||(Z + 1_n(\bar{\mu} - \bar{X})^\top)\hat{V}||_2}{\lambda}] \\ &\lesssim \mathbb{E}[||\hat{V} - VH||_F] + \mathbb{E}[\frac{||(Z + 1_n(\bar{\mu} - \bar{X})^\top)\hat{V}||_2}{\lambda}] \\ &\leq \mathbb{E}[||\hat{V} - VH||_F] + L_p\rho_2(L, M)[\frac{1 + \sqrt{\frac{s'}{n}}}{||\kappa||}], \end{aligned}$$

where $L_p$ denotes a poly-log $p$ term.

Recall that $\lambda \asymp ||\kappa||^2/\rho_2(L,M)$ (by Lemma 2.1 in [6]), $s' = p^{1-v} + p^{1-q} + \log p$ and $\rho_2(L,M) \lesssim ||\kappa||^2$, then

$$\mathbb{E}[||\hat{U} - UH||_F] \lesssim \sqrt{\left(\frac{||\kappa||^2\rho_2(L,M) + \rho_2^2(L,M)}{n||\kappa||^4}\right) \cdot (s'L_p + p^{(v-q)_+}||\kappa||^2\rho_1(L,M))) \wedge K} + L_p err_p$$

$$\leq L_p err_p.$$

Using a similar argument to the one given in the proof of Theorem 2.2 in [6] and applying the $k$-means method to $\hat{U}$ lead to a matched clustering error rate:

$$\mathbb{E}[\frac{1}{n}\text{Hamm}_p^*(\hat{y}^{\text{SPCA}}, y)] \leq L_p err_p.$$

This clustering error rate matches the rate given in Theorem 2.2 of [6].

As discussed in [4], the initialization $\hat{V}_0$ in Algorithm 1 needs to satisfy

$$|\text{supp}(\hat{V}_0)| \leq s' \quad \text{and} \quad \sigma_r(\hat{V}_0^\top V) \geq 1/2, \tag{3}$$

where $s'$ is defined in Theorem 1.

The diagonal thresholding method in the initialization procedure in [4] is designed specifically for the special case where $\Sigma = I$. In this case, (3) holds for the initialization procedure in [4] when the diagonal thresholding method is applied to the normalized data matrix $W = X - 1_n \bar{X}^\top$, where $\bar{X} = \frac{1}{n}\sum_{i=1}^n X_i$, and $1_n \in \mathbb{R}^n$ is the $n$-dimensional vector with elements equal to 1. However, the performance of the diagonal thresholding method is not guaranteed when $\Sigma$ is a general diagonal covariance matrix as considered in [6]. We replace this feature selection step by the PCA-1 step in the IF-PCA procedure, and denote the corresponding initial estimator as $\hat{V}_0$. The following theorem shows that (3) holds for $\hat{V}_0$.

**Theorem 2** *Under the conditions of Theorem 1, and suppose* $||\kappa||_\infty \to 0$. *The initial estimator $\hat{V}_0$ is the left singular vectors on $W^{\hat{S}}$ with $\hat{S}$ being the set of features selected by the PCA-1 procedure. With probability at least $1 - o(p^{-2}) - o(n^{-1})$,*

$$|\text{supp}(\hat{V}_0)| \leq s' \quad \text{and} \quad \sigma_r(\hat{V}_0^\top V) \geq 1/2.$$

*Proof:* For simplicity, we assume $\hat{S}$ and $Z$ are independent. (We can achieve this by sample splitting, or avoid this assumption by the similar argument in [6]). Note that (C.61) in [6] implies $|\text{supp}(\hat{V}_0)| \leq s'$. We thus focus on the second inequality in (3).

According to (2.18) in [6],

$$W^{\hat{S}} = LM\Lambda + L(M^{\hat{S}} - M)\Lambda + (Z\Sigma^{-1/2} + Z\Sigma^{-1/2}(\Lambda - I) + R)^{\hat{S}}$$
$$= UDV^\top + L(M - M^{\hat{S}})\Lambda + (Z\Sigma^{-1/2})^{\hat{S}} + (Z\Sigma^{-1/2}(\Lambda - I) + R)^{\hat{S}}$$
$$:= S + E_1 + E_2^{\hat{S}} + E_3^{\hat{S}}.$$

This follows

$$\begin{aligned}(W^{\hat{S}})^\top W^{\hat{S}} &= (S + E_1 + E_2^{\hat{S}} + E_3^{\hat{S}})^\top (S + E_1 + E_2^{\hat{S}} + E_3^{\hat{S}}) \\ &= S^\top S + (E_1 + E_2^{\hat{S}} + E_3^{\hat{S}})^\top S + S^\top (E_1 + E_2^{\hat{S}} + E_3^{\hat{S}}) \\ &\quad + (E_1 + E_2^{\hat{S}} + E_3^{\hat{S}})^\top (E_1 + E_2^{\hat{S}} + E_3^{\hat{S}}) \\ &= VD^2 V^\top + (E_1 + E_2^{\hat{S}} + E_3^{\hat{S}})^\top S + S^\top (E_1 + E_2^{\hat{S}} + E_3^{\hat{S}}) \\ &\quad + (E_1 + E_2^{\hat{S}} + E_3^{\hat{S}})^\top (E_1 + E_2^{\hat{S}} + E_3^{\hat{S}}) \\ &= V(D^2 + nI_p^{\hat{S}})V^\top + (E_1 + E_2^{\hat{S}} + E_3^{\hat{S}})^\top S + S^\top (E_1 + E_2^{\hat{S}} + E_3^{\hat{S}}) \\ &\quad + (E_1 + E_3^{\hat{S}})^\top (E_1 + E_3^{\hat{S}}) + (E_2^{\hat{S}})^\top (E_1 + E_3^{\hat{S}}) + (E_1 + E_3^{\hat{S}})^\top E_2^{\hat{S}} + \big((E_2^{\hat{S}})^\top E_2^{\hat{S}} - nI_p^{\hat{S}}\big). \end{aligned}$$

The rest of the proof is divided into three steps.

**Step 1.** Bounds for $||E_1||_2, ||E_2^{\hat{S}}||_2$, and $||E_3^{\hat{S}}||_2$.

Lemma 2.2 in [6] yields that with probability at least $1 - o(p^{-2})$,

$$||E_1||_2 \lesssim ||\kappa||\sqrt{n}[s^{-1/2}\sqrt{\rho_1(L,M)}\sqrt{\log p} + p^{-[(\sqrt{r}-\sqrt{q})_+]^2/2K}],$$

and it follows from the Bai-Yin law that with probability at least $1 - 2e^{-n}$,

$$||E_2^{\hat{S}}||_2 \lesssim \sqrt{n} + \sqrt{s'}.$$

In addition, by (C.70) in [6], with probability at least $1 - o(p^{-3})$,

$$||R^{\hat{S}}||_2 \lesssim [\sqrt{\log p} + \sqrt{s'\log p} + ||\kappa||s^{-1/2}\sqrt{s'\rho_1(L,M)\log p}],$$

and

$$||Z\Sigma^{-1/2}(\Lambda - I)||_2 \leq ||Z\Sigma^{-1/2}||_2 \cdot ||\Lambda - I||_2 \leq \sqrt{n}||\kappa||_\infty.$$

Combining these two inequalities leads to

$$||E_3^{\hat{S}}||_2 \leq [\sqrt{\log p} + \sqrt{s'\log p} + ||\kappa||s^{-1/2}\sqrt{s'\rho_1(L,M)\log p}] + \sqrt{n}||\kappa||_\infty.$$

**Step 2.** $||S^\top E_2^{\hat{S}}||_2 \leq \sqrt{s}||S||_2$ and $||(E_2^{\hat{S}})^\top E_2^{\hat{S}} - nI_p^{\hat{S}}|| \leq n \cdot \sqrt{\frac{s}{n}} = \sqrt{ns}$.

Let $\tilde{E}_i$ be the $i$-th column of $E_2^{\hat{S}}$. Since $E_2^{\hat{S}} = (Z\Sigma^{-1/2})^{\hat{S}}$ has i.i.d. $N(0,1)$ entries, $Y_i = U^\top \tilde{E}_i \sim N_{K-1}(0, I_{K-1})$. Let $Y \in \mathbb{R}^{(K-1)\times s'}$ with the $i$-th column being $Y_i$. By the Bai-Yin law, with probability at least $1 - e^{-s'}$,

$$\|Y\|_2 \lesssim \sqrt{K} + \sqrt{s'}.$$

This implies
$$\|S^\top E_2^{\hat{S}}\|_2 \leq \|VDU^\top E_2^{\hat{S}}\|_2 \leq \|D\|_2 \|Y\|_2 \lesssim \sqrt{s'} \|S\|_2.$$

In addition, since the entries of $E_2^{\hat{S}}$ are i.i.d. $N(0,1)$, then according to [8],

$$\|(E_2^{\hat{S}})^\top E_2^{\hat{S}} - nI_p^{\hat{S}}\| \leq n \cdot \sqrt{\frac{s}{n}} = \sqrt{ns}.$$

**Step 3.** $\sigma_r(\hat{V}_0^\top V) \geq \frac{1}{2}$.

By Davis-Kahan sin-theta Theorem, there exists $H \in O(K-1, K-1)$ such that

$$\|\hat{V}_0 - VH\|_F \leq \frac{\frac{n\|\kappa\|}{\sqrt{s}}\|\kappa\|/\rho_2(L,M)\sqrt{\rho_1(L,M)}\sqrt{\log p}}{n\|\kappa\|^2/\rho_2^2(L,M)}$$
$$\leq \frac{\sqrt{\rho_1(L,M)}\sqrt{\log p}}{\sqrt{s}/\rho_2(L,M)}$$
$$\leq p^{-C} \to 0,$$

where the last inequality follows from (2.15) in [6].

In conclusion, if we let $\Delta = \hat{V}_0 - VH$, then $\|\Delta\|_2 \leq p^{-C}$. This indicates that when $p$ is sufficiently large, the $r$-th largest singular value of $\hat{V}^\top V$ satisfies

$$\sigma_r(\hat{V}_0^\top V) = \sigma_r((VH + \Delta)^\top V) = \sigma_r(H^\top V^\top V + \Delta^\top V)$$
$$\geq 1 - \|\Delta^\top V\|_2 \geq 1/2.$$

We now compare the numerical performance of the SPCA method with the IF-PCA method in the same settings considered in the simulation section of [6] with $p = 4000$ and $n = 145$. $\Sigma$ is nearly an identity in their settings, so we use the initial estimator in [4] with the data matrix normalized by centering only, and the simulation results suggest a robust performance of this SPCA clustering method.

Recall that $r$ indicates the strength of the signal, and the sparsity is $p^{1-v}$. We calculate the clustering error rates of IF-PCA and SPCA for the combinations $\{r, v\} = \{0.25, 0.35, 0.5, 0.65\} \times \{0.6, 0.7, 0.8\}$ with $K = 2$. The simulation results are summarized in Table 1. The results show that the clustering method based on the SPCA procedure introduced in [4] outperforms IF-PCA in most cases. The numerical results are consistent with the theoretical results given in Theorem 1.

Table 1: Clustering error rate of IF-PCA and SPCA for each $\{r, v\}$ combination.

| $r$ | Method | $v$ 0.8 | 0.7 | 0.6 |
|---|---|---|---|---|
| 0.25 | IF-PCA | 0.4541 | 0.4045 | 0.1369 |
|  | SPCA | **0.4234** | **0.3396** | **0.0124** |
| 0.35 | IF-PCA | 0.4417 | 0.3917 | 0.1000 |
|  | SPCA | **0.4176** | **0.2200** | **0.0031** |
| 0.5 | IF-PCA | 0.4217 | 0.2683 | 0.0245 |
|  | SPCA | 0.4252 | **0.1834** | **0.0013** |
| 0.65 | IF-PCA | 0.4072 | 0.2290 | 0.0452 |
|  | SPCA | 0.4266 | **0.0959** | **0.0017** |

In addition, we compare the IF-PCA and SPCA clustering methods in the six gene microarray data sets considered in [6]. In this comparison, the tuning parameters in SPCA are fixed at $\alpha = 1$ for the initialization, and $\beta = 1, \delta = 0.2$ for all six cases. Under this setting, the results given Table 2 and Figure 2 indicate that the SPCA clustering method is competitive with the IF-PCA method. We believe that an SPCA procedure with optimally tuned parameters would further improve the numerical results.

Table 2: Clustering error for 6 gene microarray data sets introduced in [6].

| Data set | $K$ | $n$ | $p$ | IF-PCA | SPCA |
|---|---|---|---|---|---|
| Brain | 5 | 42 | 5597 | 0.262 | 0.190 |
| Leukemia | 2 | 72 | 3571 | 0.069 | 0.028 |
| Lung Cancer (1) | 2 | 181 | 12533 | 0.033 | 0.083 |
| Prostate Cancer | 2 | 102 | 6033 | 0.382 | 0.422 |
| SRBCT | 4 | 63 | 2308 | 0.444 | 0.508 |
| Lymphoma | 3 | 62 | 4026 | 0.065 | 0.016 |

The above theoretical and numerical analyses indicate that the SPCA based clustering method has similar performance as that of the IF-PCA method. It is important to note that both the IF-PCA and SPCA methods require the assumption that $\Sigma$ is a diagonal matrix. We will discuss this assumption in the next section.

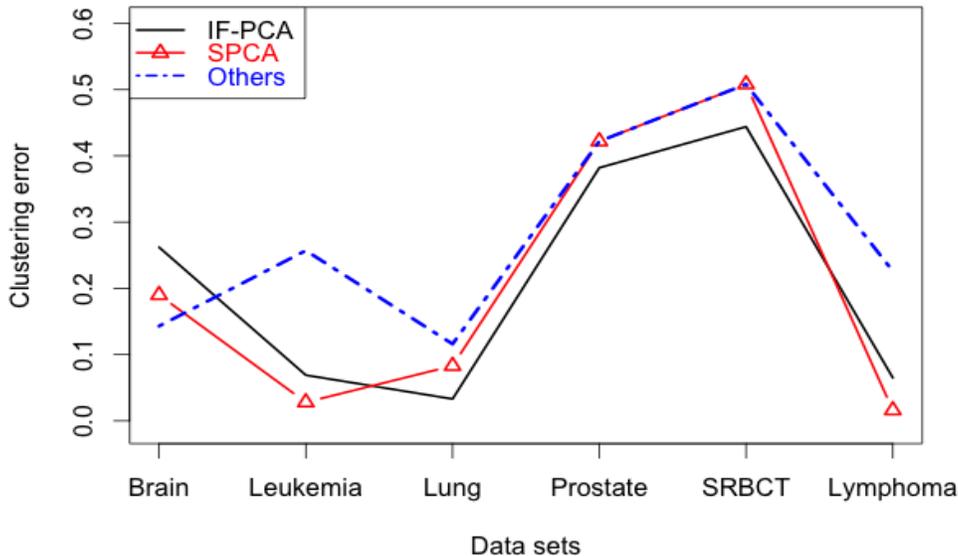

Figure 1: Comparison of the clustering errors of the IF-PCA and SPCA methods for the six gene microarray datasets. 'Others' stands for the minimum of the error rates of all other methods (except the IF-PCA method) in [6].

## 3 General Covariance Structure

[6] focuses on the special case where the common covariance matrix $\Sigma$ among the mixture components is diagonal. This assumption is quite restrictive but it is essential for the success of the IF-PCA procedure. We now consider the dependent case with a general covariance matrix $\Sigma$ that is not necessarily diagonal and demonstrate that the screening step may adversely affect the efficiency of the subsequent clustering method, even if all the "useless features" are correctly screened out. In this sense, the IF-PCA procedure is specifically design for the case of diagonal $\Sigma$.

Let us first consider an oracle setting where the number of mixture components $K = 2$ (the case where $K \geq 3$ can be similarly considered [3]), and the true parameters $\bar{\mu}$, $\mu_k (k = 1, ..., K)$, and $\Sigma$ are known. We further assume $X_i | y_i = k \sim N_p(\bar{\mu} + \mu_k, \Sigma)$, and $P(y_i = k) = \delta_k$. The goal is to cluster the sample data given these true parameters. In this case, the optimal clustering procedure is Fisher's linear discriminant rule:

$$\psi(Z) = 1 + I_{\{(Z-\mu)'\Sigma^{-1}\Delta \geq \log(\delta_1/\delta_2)\}},$$

where $\mu = \bar{\mu} + (\mu_1 + \mu_2)/2$, $\Delta = \mu_1 - \mu_2$, and this rule labels the data point $X_i \in \mathbb{R}^p$ to class $\psi(X_i)$. This classifier is the Bayes rule with the prior probabilities $\delta_1$ and $\delta_2$ for classes 1 and 2 respectively, and is thus optimal in such an ideal setting.

The misclassification error rate of Fisher's rule [1] is give by
$$R_{\text{Fisher}} = 1 - \Phi(\sqrt{\Delta'\Sigma^{-1}\Delta}),$$
which is the best possible performance when all the parameters are known in advance.

To see that the screening step, which is solely based on the means, is not always desirable, write
$$\Delta = \begin{pmatrix} \Delta_1 \\ \Delta_2 \end{pmatrix} = \begin{pmatrix} \mu_{11} - \mu_{21} \\ \mu_{21} - \mu_{22} \end{pmatrix} \text{ and } \Sigma = \begin{pmatrix} \Sigma_{11} & \Sigma_{12} \\ \Sigma_{21} & \Sigma_{22} \end{pmatrix},$$
where $\Delta_1$ is a $s$-dimensional vector, $\Sigma_{11}$ is $s \times s$, $\Sigma_{12}$ is $s \times (p-s)$ and $\Sigma_{22}$ is $(p-s) \times (p-s)$. Let $\Delta_1 \neq 0$ and $\Delta_2 = 0$. That is, $\Delta_1$ contains all the useful features and $\Delta_2$ corresponds to the set of all "useless features". Suppose we correctly screen out all the $p - s$ "useless features" and clustering the data based on the first $s$ features. The next inequality shows that Fisher's rule in the oracle setting based on all the features outperforms Fisher's rule based only on the useful features:
$$\Delta'\Sigma^{-1}\Delta = \Delta_1'\Sigma_{11}^{-1}\Delta_1 + (\Delta_2 - \Sigma_{22}^{-1}\Delta_1)\Sigma_{22\cdot 1}^{-1}(\Delta_2 - \Sigma_{22}^{-1}\Delta_1) \geq \Delta_1'\Sigma_{11}^{-1}\Delta_1,$$
where the inequality follows from the fact that $\Sigma_{22\cdot 1} = \Sigma_{22} - \Sigma_{21}\Sigma_{11}^{-1}\Sigma_{21} \geq 0$. This inequality implies
$$\Phi(\Delta'\Sigma^{-1}\Delta) - \Phi(\Delta_1'\Sigma_{11}^{-1}\Delta_1) \geq 0.$$

We now consider the data-driven IF-PCA procedure. Denote $\hat{y}^{\text{IF}-\text{all}}$ and $\hat{y}^{\text{IF}-\text{u}}$ as the IF-PCA clustering method based on all features and useful features respectively, and similarly for $\hat{y}^{\text{Fisher}-\text{all}}$ and $\hat{y}^{\text{Fisher}-\text{u}}$. Recall the clustering error for the IF-PCA method defined in [6] is
$$L(\hat{y}^{\text{IF}}, y) = \frac{1}{n} \min_\pi \{\sum_{i=1}^n I(\hat{y}^{\text{IF}} \neq \pi(y_i))\},$$
where $\pi$ is any permutation of $\{1, 2, ..., K\}$.

According to the optimality of Fisher's rule,
$$\mathbb{E}[L(\hat{y}^{\text{IF}-\text{u}}, y)] = \mathbb{E}[\frac{1}{n}\sum_{i=1}^n I(\hat{y}^{\text{IF}-\text{u}} \neq y_i)] = \mathbb{E}[I(\hat{y}^{\text{IF}-\text{u}} \neq y_i)]$$
$$\geq \mathbb{E}[I(\hat{y}^{\text{Fisher}-\text{u}} \neq y_i)]$$
$$= \mathbb{E}[I(\hat{y}^{\text{Fisher}-\text{all}} \neq y_i)] - (\Phi(\Delta'\Sigma^{-1}\Delta) - \Phi(\Delta_1'\Sigma_{11}^{-1}\Delta_1))$$
$$= \mathbb{E}[I(\hat{y}^{\text{IF}-\text{all}} \neq y_i)] + \text{Err} - (\Phi(\Delta'\Sigma^{-1}\Delta) - \Phi(\Delta_1'\Sigma_{11}^{-1}\Delta_1)),$$
where $\text{Err} = (\mathbb{E}[I(\hat{y}^{\text{Fisher}-\text{all}} \neq \pi(y_i))] - \mathbb{E}[I(\hat{y}^{\text{IF}-\text{all}} \neq \pi(y_i))])$ is the statistical error which goes to zero as the sample size $n$ goes to infinity. Therefore, there exists $\mu_1, \mu_2, \Sigma, n$, such that $\text{Err} - (\Phi(\Delta'\Sigma^{-1}\Delta) - \Phi(\Delta_1'\Sigma_{11}^{-1}\Delta_1)) < 0$, and then
$$\mathbb{E}[L(\hat{y}^{\text{IF}-\text{u}}, y)] > \mathbb{E}[L(\hat{y}^{\text{IF}-\text{all}}, y)].$$

The above discussion suggests that, when $\Sigma_{12} \neq 0$, screening based on the means alone may in fact increase the clustering error even if it identifies all the "useless features". Whether or not a feature is useless not only depends on the difference in the two means but also depends on the covariance structure. The optimality achieved by IF-PCA in the independent case, where $\Sigma$ is diagonal, thus no longer holds in the general case due to the screening procedure.

It is an interesting future research project to study if the IF-PCA method can be generalized to achieve good clustering results without the diagonality assumption on $\Sigma$. It appears that a good screening step based on both the means and covariances is essential for the success of such a two-step procedure.